
%
%
\input harvmac

\def\CTPa{\it Center for Theoretical Physics, Department of Physics,
      Texas A\&M University}
\def\CTPb{\it College Station, TX 77843-4242, USA}
\def\HARCa{\it Astroparticle Physics Group,
Houston Advanced Research Center (HARC)}
\def\HARCb{\it The Woodlands, TX 77381, USA}

\def\ie{\hbox{\it i.e.}}     
\def\eg{\hbox{\it e.g.}}

\def\coeff#1#2{{\textstyle{#1\over #2}}}

\catcode`\@=11 

\def\lsim{\mathrel{\mathpalette\@versim<}}
\def\gsim{\mathrel{\mathpalette\@versim>}}
\def\@versim#1#2{\vcenter{\offinterlineskip
    \ialign{$\m@th#1\hfil##\hfil$\crcr#2\crcr\sim\crcr } }}
\def\boxit#1{\vbox{\hrule\hbox{\vrule\kern3pt
      \vbox{\kern3pt#1\kern3pt}\kern3pt\vrule}\hrule}}

\def\t1{{\tilde 1}}

\def\F{\widetilde F}

\def\JL{J. L. Lopez}
\def\DVN{D. V. Nanopoulos}

\def\mol{\rm Mo\hskip-4.5pt/\hskip1pt l}

\def\GeV{\,{\rm GeV}}

\def\wt{\widetilde}
\def\ep{\epsilon}

\def\NPB#1#2#3{Nucl. Phys. B {\bf#1} (19#2) #3}
\def\PLB#1#2#3{Phys. Lett. B {\bf#1} (19#2) #3}

\def\PRD#1#2#3{Phys. Rev. D {\bf#1} (19#2) #3}
\def\PRL#1#2#3{Phys. Rev. Lett. {\bf#1} (19#2) #3}
\def\PRT#1#2#3{Phys. Rep. {\bf#1} (19#2) #3}
\def\MODA#1#2#3{Mod. Phys. Lett. A {\bf#1} (19#2) #3}

\def\TAMU#1{Texas A \& M University preprint CTP-TAMU-#1}
\def\ARAA#1#2#3{Ann. Rev. Astron. Astrophys. {\bf#1} (19#2) #3}
\def\ARNP#1#2#3{Ann. Rev. Nucl. Part. Sci. {\bf#1} (19#2) #3}

\nref\dmreview{For reviews on dark matter and its detection, see \eg,
V. Trimble, \ARAA{25}{87}{425};
J. R. Primack, B. Sadoulet, and D. Seckel, \ARNP{38}{88}{751}; D. O. Caldwell,
\MODA{5}{90}{1543}; P. F. Smith and J. D. Lewin, \PRT{187}{90}{203}.}
\nref\olddm{S. Weinberg, \PRL{50}{83}{387}; H. Goldberg, \PRL{50}{83}{1419};
L. M. Krauss, \NPB{227}{83}{556}.}
\nref\EHNOS{J. Ellis, J. S. Hagelin, D. V. Nanopoulos, K. A. Olive
and M. Srednicki, \NPB{238}{84}{453}.}
\nref\cdmreview{For a recent review see M. Davis, G. Efstathiou, C. S. Frenk,
and S. D. M. White, {\it Nature} {\bf356} (1992) 489.}
\nref\KT{See \eg, E. Kolb and M. Turner, {\it The Early Universe}
(Addison-Wesley, Redwood City, 1990).}
\nref\olivereview{For recent reviews see \eg, K. Olive, \PRT{190}{90}{307};
D. Goldwirth and T. Piran, \PRT{214}{92}{223}.}
\nref\minimaldm{J. Ellis, J. Hagelin, and D. V. Nanopoulos, \PLB{159}{85}{26};
K. Griest, \PRD{38}{88}{2357}; \PRD{39}{89}{3802}(E);
K. A. Olive and M. Srednicki, \PLB{230}{89}{78} and \NPB{355}{91}{208};
K. Griest, M. Kamionkowski, and M. S. Turner, \PRD{41}{90}{3565};
J. Ellis, L. Roszkowski, and Z. Lalak, \PLB{245}{90}{545};
L. Roszkowski, \PLB{262}{91}{59}; K. Griest and L. Roszkowski,
\PRD{46}{92}{3309}; L. Roszkowski, \PLB{278}{92}{147};
J. McDonald, K. A. Olive, and M. Srednicki, \PLB{283}{92}{80}; S. Mizuta
and M. Yamaguchi, \PLB{298}{93}{120}; S. Mizuta, D. Ng, M. Yamaguchi,
TU-410 (August 1992).}
\nref\nonminimaldm{R. Flores, K. A. Olive and D. Thomas, \PLB{245}{90}{509};
K. Olive and D. Thomas, \NPB{355}{91}{192}; S. Abel, S. Sarkar, and
I. Whittingham, OUTP-92-10P (September 1992).}
\nref\without{M.M. Nojiri, \PLB{261}{91}{76}; \JL, \DVN, and K. Yuan,
\PLB{267}{91}{219}; J. Ellis and L. Roszkowski, \PLB{283}{92}{252}.}
\nref\LNYa{\JL, \DVN, and K. Yuan, \NPB{370}{92}{445}.}
\nref\with{J. Ellis and F. Zwirner, \NPB{338}{90}{317}; S. Kelley, \JL, \DVN,
H. Pois, and K. Yuan, \PLB{273}{91}{423}; M. Kawasaki and S. Mizuta,
\PRD{46}{92}{1634}; M. Dress and M. M. Nojiri, \PRD{47}{93}{376}; S. Kelley,
\JL, \DVN, H. Pois, and K. Yuan, \PRD{47}{93}{2461};
R. Roberts and L. Roszkowski, RAL-93-003 and UM-TH-33/92.}
\nref\LNZa{\JL, \DVN, and A. Zichichi, \PLB{291}{92}{255}.}
\nref\LNP{\JL, \DVN, and H. Pois, \PRD{47}{93}{2468}.}
\nref\SWO{M. Srednicki, R. Watkins, and K. Olive, \NPB{310}{88}{693}.}
\nref\GG{P. Gondolo and G. Gelmini, \NPB{360}{91}{145}.}
\nref\GS{K. Griest and D. Seckel, \PRD{43}{91}{3191}.}
\nref\LN{For a review see, A. B. Lahanas and D. V. Nanopoulos,
\PRT{145}{87}{1}.}
\nref\aspects{See \eg, S. Kelley, \JL, \DVN, H. Pois, and K. Yuan, \TAMU{16/92}
and CERN-TH.6498/92 (to appear in Nucl. Phys. B).}
\nref\Dickreview{For reviews see: R. Arnowitt and P. Nath, {\it Applied N=1
Supergravity} (World Scientific, Singapore 1983);
H. P. Nilles, \PRT{110}{84}{1}; and Ref. \LN.}
\nref\AN{M. Matsumoto, J. Arafune, H. Tanaka, and K. Shiraishi,
\PRD{46}{92}{3966}; R. Arnowitt and P. Nath, \PRL{69}{92}{725}; P. Nath and
R. Arnowitt, \PLB{287}{92}{89} and \PLB{289}{92}{368}.}
\nref\LNPZ{\JL, \DVN, H. Pois, and A. Zichichi, \PLB{299}{93}{262}.}
\nref\LNPWZh{\JL, \DVN, H. Pois, X. Wang and A. Zichichi, \TAMU{05/93}
(to appear in Phys. Lett. B).}
\nref\ANcosmo{R. Arnowitt and P. Nath, \PLB{299}{93}{58} and Erratum;
P. Nath and R. Arnowitt, NUB-TH-3056/92, CTP-TAMU-66/92 (revised).}

\nfig\I{The thermal average $\vev{\sigma v_{\mol}}$ (in arbitrary units) as a
function of $y_z=(M_Z/2m_\chi)^2$ for a $Z$-pole dominance situation. The exact
and traditional series expansion results are shown. Note that the second-order
approximation gives negative thermal averaged annihilation cross sections
right above the pole.}
\nfig\II{Same as Fig. 1 but for an $h$-pole dominance situation.}
\nfig\III{The ratios $\vev{\sigma v_{\mol}}_{exact}/\vev{\sigma
v_{\mol}}_{approx}$ for the two approximate series expansions (first-order:
dashed; second-order: solid) and for the two
sample poles considered. The peak in the $a+bx+cx^2$ line is due to the series
expansion changing sign at $y=0.75$, giving negative (non-sensical) results for
$0.75<y\lsim1$.  The bottom row plots show the slow convergence
not-so-near the poles to the exact result.}
\nfig\IV{The integrand in the thermal average (Eq. (2.9)) as a function of
$y$ for two representative values of $y_R$ and the two poles considered.
The $w(y)$ function is used in exact form (solid lines) and Taylor-expanded
form up to first (dotted lines) and second (dashed lines) derivatives. Note
that the expansions miss the poles completely, although these are killed-off
when sufficiently away from the poles.}
\nfig\V{The calculated values of the neutralino relic density in the minimal
$SU(5)$ supergravity model as a function of $m_\chi$ using the exact thermal
average procedure. Note that the pole structure near the $Z$-pole is
broader and shallower than that shown in Fig. 6 where the series approximation
to the thermal average has been used.}
\nfig\VI{Same as Fig. 5 but using the first order series approximation to
the thermal average ($\vev{\sigma v_{\mol}}=a+bx$).}
\nfig\VII{Those points in parameter space where $\Omega_\chi h^2_0<1$ with
the thermal average computed exactly. These results do not differ qualitatively
from those shown in Fig. 2 in Ref. \LNPZ, where the series approximation to
the thermal avarage was used.}
\nfig\VIII{Those points in parameter space where
$\Omega_\chi^{exact}/\Omega_\chi^{approx}>1$, which according to Fig. 3
should occur near and above the poles. The dashed lines indicate the $Z$-
and $h$-poles.}
\nfig\VIII{Those points in parameter space where
$\Omega_\chi^{exact}/\Omega_\chi^{approx}<0.5$, which according to Fig. 3
should occur below the poles. The dashed lines indicate the $Z$-
and $h$-poles, as well as the $\chi\chi\to hh$ threshold.}

\Title{
\vbox{\baselineskip12pt\hbox{CTP--TAMU--14/93}\hbox{ACT--4/93}}}
{\vbox{\centerline{Accurate Neutralino Relic Density Computations}
\vskip2pt\centerline{in Supergravity Models}}}
\centerline{JORGE~L.~LOPEZ, D.~V.~NANOPOULOS, and KAJIA YUAN}
\smallskip
\centerline{\CTPa}
\centerline{\CTPb}
\centerline{and}
\centerline{\HARCa}
\centerline{\HARCb}
\vskip .1in
\centerline{ABSTRACT}
We  investigate the question of the proper thermal averaging of
neutralino annihilation amplitudes which possess poles and thresholds, as
they impact on the calculated neutralino relic density and therefore on
the cosmological viability of supersymmetric models. We focus on two typical
resonances, namely the $Z$ boson and the lightest Higgs boson ($h$). In the
context of supergravity models with radiative electroweak
symmetry breaking, an exploration of the whole parameter space of the model
is possible and the overall relevance of these sophisticated analyses can
be ascertained. As an example we chose the minimal $SU(5)$ supergravity
model since the presence of such poles is essential to obtain a cosmologically
acceptable model. We find that the proper thermal averaging is important for
individual points in parameter space and that the fraction of cosmologically
acceptable points  is increased somewhat by the accurate procedure. However,
qualitatively the new set of points is very similar to that obtained previously
using the usual series approximations to the thermal average. We conclude that
all phenomenological analyses based on the previously determined cosmologically
allowed set remain valid.
\Date{February, 1993}

\newsec{Introduction}
Much speculation has gone on for some time about the nature of the ``observed"
astrophysical dark matter in the Universe \dmreview. The lightest
supersymmetric particle -- the lightest neutralino -- is a prime candidate for
a cold dark matter relic \refs{\olddm,\EHNOS},
and as such it would constitute an essential ingredient in contemporary
structure formation ideas \cdmreview. A consistency check for a possible dark
matter candidate $\chi$ in the Big-Bang cosmology is provided by an independent
lower bound on the age of the Universe. The present relic abundance
$\Omega_\chi=\rho_\chi/\rho_{crit}$ must be bounded above by $\Omega_\chi
h^2_0<1$ \KT, where $0.5<h_0<1$ is the Hubble parameter in units of
$100\,{\rm km}\,{\rm s}^{-1}\,{\rm Mpc}^{-1}$. Weaker bounds follow from direct
astrophysical determinations of $\Omega_0$. Also, in  inflationary cosmology
$\Omega_0\equiv1$ \olivereview\ and the observational bound is also satisfied.

In the specific case of the lightest neutralino, the computation of
$\Omega_\chi$ has been attempted in several supersymmetric (minimal
\refs{\olddm,\EHNOS,\minimaldm}, non-minimal \nonminimaldm, supergravity
without \refs{\without,\LNYa}, and with \refs{\with,\LNZa,\LNP} radiative
electroweak breaking) models to varying degrees of approximation for nearly a
decade. The details of the particle physics model come into play mainly in the
calculation of the total annihilation amplitude $\chi\chi\to all$ and its
thermal average. Under normal circumstances, thermal considerations allow one
to conclude that center-of-mass energies close to the minimal one
($\sqrt{s}=2m_\chi$) are the most likely ones \EHNOS, and that a rapidly
converging series expansion around this point should suffice \SWO. For
two-body final states with masses $m_{1,2}$ to be ``open" (in the
non-relativistic limit) it is necessary that $m_1+m_2\le2m_\chi$. Since $\chi$
is the lightest supersymmetric particle, only non-supersymmetric final states
contribute (\ie, $q\bar q,l^+l^-,W^+W^-,
ZZ,hh,hA,hH,Zh,ZA,\cdots$, where $h,A,H$ are the supersymmetric Higgs bosons).

Such series expansions, however, have been noted to fail badly when
$s$-channel resonances and/or new-channel thresholds are present in the
annihilation amplitude \refs{\GG,\GS}. The discussion in these references
provided a solid ground for tackling such problems, but their practical
applications were only briefly explored. On the other hand, supergravity models
with radiative electroweak symmetry breaking \LN\ provide a fertile testing
ground for such sophisticated techniques, since the complete mass spectrum and
couplings of the model can be specified in terms of very few parameters
\aspects. In particular, the occurence of poles and thresholds in the
annihilation amplitude can be studied over the whole parameter space of these
models, in order to determine whether the more accurate results provided by
these techniques change significantly the cosmologically allowed region
of parameter space or not. These special considerations play a very important
role in the cosmological viability of the longest-standing GUT model, namely
the minimal $SU(5)$ supergravity model \Dickreview. It has been recently
pointed out \refs{\LNZa,\LNP} that, because the stringent proton decay
constraints on this model force the most efficient annihilation channel
mediators to be very heavy, the neutralino relic density is small enough
basically only near the lightest Higgs ($h$) and $Z$-boson resonances, \ie,
for $m_\chi\approx{1\over2}m_{h,Z}$. Thus, a more accurate treatment, as
described in Refs. \refs{\GG,\GS}, appears mandatory in this case.

The purpose of this paper is twofold. Firstly, we
present a quantitative discussion of the accurate thermal averaging necessary
in the context of this class of models, and explicitly show the breakdown of
the usual series expansion near the poles and its not-so-good accuracy
not-so-near the poles. Secondly, we perform a complete recalculation of the
relic density in the minimal $SU(5)$ supergravity model, and show the effects
of the more accurate treatment and the explicit role of the $s$-channel
resonances ($h,Z$) and the $\chi\chi\to hh$ threshold. We find that even though
the re-computed values of the relic density are shifted  relative to our
previous (less accurate) results, the overall fraction of the parameter space
which is cosmologically allowed is not qualitatively changed, and thus all
predictions based on the previously determined cosmologically allowed set
remain valid.

\newsec{The thermal average}
The neutralino relic density is given by the expression \SWO
\eqn\bA{\Omega_\chi h^2_0=1.555\times10^8(m_\chi/\GeV)h(0)q(0),}
where $h(0)=3.91$ is the effective number of
entropy degrees of freedom today, and $q(0)$ is obtained by solving the
Boltzmann equation for $q\equiv n/(T^3h(T))$, with $n$ the actual number
density of $\chi$ particles at temperature $T$. This equation is given by
\eqn\A{{dq\over dx}=\lambda(x)(q^2-q^2_0)(x)}
with $x=T/m_\chi$, $q_0$ the analog of $q$ but with the $\chi$ particles in
thermal equilibrium, and
\eqn\B{\lambda(x)=\left(\coeff{4}{45}\pi^3G_N\right)^{-1/2}
{m_\chi\over\sqrt{g(T)}}\,[h(T)+\coeff{1}{3}m_\chi xh'(T)]\,\vev{\sigma
v_{\mol}},}
where $G_N$ is the gravitational constant and $g(T)$ the effective number of
energy density degrees of freedom. For a detailed discussion of how to
evaluate all the terms appearing in the Boltzmann equation and how to solve the
equation itself, see \eg, the Appendix in Ref. \LNYa. The novelty in the
present discussion is in the evaluation of the thermal average factor
$\vev{\sigma v_{\mol}}$, where $v_{\mol}$ is the M\o ller velocity \GG. The
general expression for this quantity is \GG
\eqn\C{\vev{\sigma v_{\mol}}={1\over8m^4_\chi T K^2_2(x^{-1})}
\int_{4m^2_\chi}^\infty \sigma\cdot(s-4m^2_\chi)\sqrt{s}\,K_1(\sqrt{s}/T)\,ds,}
where $K_i$ are the modified Bessel functions of order $i$, and $\sigma$ is
the total annihilation cross section. To make contact with previous work we
rewrite $\vev{\sigma v_{\mol}}$ in Lorentz invariant form as follows
\eqn\D{\vev{\sigma v_{\mol}}={1\over4m^5_\chi x K^2_2(x^{-1})}
\int_{4m^2_\chi}^\infty ds(s-4m^2_\chi)^{1/2}K_1(\sqrt{s}/xm_\chi)\,w(s),}
with \SWO
\eqn\E{w(s)=\coeff{1}{4}\int d{\rm LIPS}\,|{\cal A}(\chi\chi\to all)|^2.}
The usual series expansion for the thermal average follows from the observation
that only for $x\lsim0.1$ is the Boltzmann equation sensitive to the value of
$\lambda(x)$. In this regime the argument of $K_1$ in Eq. \D\ is always larger
than $\sqrt{s}/xm_\chi>2/x\gsim20$, and the Bessel function
($K_1(y)\sim\sqrt{\pi/2y}\,e^{-y},\,y\gg1$) dies away quickly with increasing
$\sqrt{s}$. Therefore a series expansion of $w(s)$ around $\sqrt{s}=2m_\chi$
should converge quickly. The resulting series of integrals can be done
analytically giving \SWO
\eqna\F
$$\eqalignno{\vev{\sigma v_{\mol}}&={1\over m^2_\chi}[w-\coeff{3}{2}(2w-w')x
+\coeff{3}{8}(16w-8w'+5w'')x^2+{\cal O}(x^3)]_{s=4m^2_\chi}&\F a\cr
&\equiv a+bx+cx^2+{\cal O}(x^3).&\F b\cr}$$

The problem with this approximation when $w(s)$ has a pole can be best seen in
a simple analytical example. Let us consider the case where an $s$-channel
resonance with mass $m_R$ and width $\Gamma_R$ dominates $w(s)$, \ie,
\eqn\G{w(y)={\cal C}{y(y-1)\over(y-y_R)^2+\gamma^2_R},}
where $y=s/4m^2_\chi$, $y_R=m^2_R/4m^2_\chi$, and $\gamma_R=\Gamma_Rm_R/
4m^2_\chi=(\Gamma_R/m_R)y_R$. This form in fact applies to $\chi\chi\to h,Z\to
f\bar f$ when $m_f=0$, with ${\cal C}$ some dimensionless function of the
couplings (see Eqs. A.2 and A.3). We then obtain
\eqna\H
$$\eqalignno{\vev{\sigma v_{\mol}}&={2{\cal C}\over m^2_\chi xK^2_2(x^{-1})}
\int_1^\infty dy{y(y-1)\over
(y-y_R)^2+\gamma^2_R}\,\sqrt{y-1}\,K_1(2\sqrt{y}/x)\cr
&\to {2{\cal C}\over \sqrt{\pi}m^2_\chi x^{3/2}}
\int_1^\infty dy {y(y-1)\over(y-y_R)^2+\gamma^2_R}\,\sqrt{y-1\over\sqrt{y}}
\,e^{-2(\sqrt{y}-1)/x},&\H{}\cr}$$
where the limiting form holds in the $x$-regime of interest ($x\lsim0.1$).
Expanding $w(y)$ around $y=1$ and performing the integrals analytically one
arrives at the series expansion in Eq. \F{} with
\eqna\I
$$\eqalignno{a&=0,&\I a\cr
b&=\coeff{3}{2}{1\over(1-y_R)^2+\gamma^2_R},&\I b\cr
c&=\coeff{3}{4}{1\over(1-y_R)^2+\gamma^2_R}
\left[1-{10(1-y_R)\over(1-y_R)^2+\gamma^2_R}\right].&\I c\cr}$$
For illustrative purposes we have computed $\vev{\sigma v_{\mol}}$ exactly
using Eq. \H{}  and also using the  expansions in Eqs. \F{},\I{}. The results
are shown in Fig. 1 (2) (in arbitrary units) for the $Z\,(h)$-pole, where we
have taken $x=0.05$ and $\Gamma_R/m_R=0.0274\,(0.0002)$ for $R=Z\,(h)$.
Clearly, sufficiently away from the poles ($y_R\gg1$ or $y_R\ll1$) the series
expansions approach the exact result, while near the poles these are highly
innacurate. In particular, the second-order approximation ($a+bx+cx^2$) fails
badly right above the poles ($y_R<1$) since it gives negative thermal averaged
annihilation cross sections.

The degree of inaccuracy of the series espansions can be better appreciated by
examining the ratios $\vev{\sigma v_{\mol}}_{exact}/\vev{\sigma
v_{\mol}}_{approx}$, as shown in Fig. 3 for both resonances. Below the pole
($y_R>1)$ one can clearly see the quicker convergence of the higher-order
approximation. The peak in the $a+bx+cx^2$ line is due to the series expansion
changing sign at $y=0.75$, giving negative (non-sensical) results for
$0.75<y\lsim1$. The overall result is that as one approaches the poles from
below ($y_R>1)$ the exact thermal average first becomes larger than naively
expected (relic density smaller), then near the poles it becomes smaller (relic
density larger), until above the poles where it quickly approaches the naive
estimate (in first order, $a+bx$). In practice this means that the neutralino
relic density distributions as a function of the neutralino mass will show
not-as-narrow (broader) and not-as-deep (shallower) pole structures, and that
these will be asymmetric.

The bottom row in Fig. 3 shows a detail of the convergence of the series
expansions to the exact result below the pole. It is perhaps somewhat
unexpected that the expansions are  relatively innacurate not-so-near
the poles, \ie, for $y_R=2$, $m_\chi\approx0.7({1\over2}m_R)$, which for
the $Z$-pole gives $m_\chi\approx32\GeV$. The reason for this behavior can
be  understood by studying the integrand in Eq. \H{} with $w(y)$ in
exact and Taylor-expanded (around $y=1$) forms. These are shown as functions
of $y$ in Fig. 4 for both poles and representative values of $y_R$. The
solid/dotted/dashed lines correspond to the integrand evaluated using $w(y)$
in exact/up-to-first-derivates/up-to-second-derivatives forms. Clearly, the
poles at $y=y_R$ are `missed' by the expansions for $y_R\gsim1$. However, for
sufficiently high values of $y_R$ the exponential kills-off the poles
completely, although they `last longer' for the much narrower $h$-pole.
Note also the quantitative effect of the expansion up to second-derivatives
(dashed lines) relative to that up to first-derivatives (dotted lines). In
practical relic density calculations only the first derivatives of $w(y)$
are usually kept (which corresponds approximately to $\vev{\sigma v_{\mol}}
=a+bx$) and therefore good accuracy is not reached until considerably away
from the poles (from below; from above convergence is faster). Since $w(y)$
usually receives additional contributions from non-resonant channels which
are likely to overshadow the contributions from resonant channels when away
from the corresponding resonances, in practive the behavior away from the poles
is not easily distinguishable.

\newsec{Application to the minimal $SU(5)$ supergravity model}
As mentioned above, the study of the relic density of neutralinos requires
the knowledge of the total annihilation amplitude $\chi\chi\to all$. The
latter depends on the model parameters to determine all masses and couplings.
Previously \LNYa\ we have advocated the study of this problem in the context
of supergravity models with radiative electroweak symmetry breaking, since
then only a few parameters (five or less) are needed to specify the model
completely. In particular, one can explore the whole parameter space and
draw conclusions about a complete class of models. The ensuing relationships
among the various masses and couplings have been found to yield results which
depart from the conventional minimal supersymmetric standard model (MSSM) lore,
where no such relations exist. Here we study a novel aspect of these
correlations, namely the ocurrence of poles and thresholds in the annihilation
amplitude. We choose to work with the minimal $SU(5)$ supergravity model
\Dickreview\ since its five-dimensional parameter space is strongly constrained
by the proton lifetime and a sufficiently small neutralino relic density. In
fact, the latter is cosmologically acceptable only because of enhancements in
the neutralino annihilation amplitude near $s$-channel $Z$ and $h$ resonances.
Our purpose here is to determine whether an accurate computation of the
thermal avarage changes significantly the previously (inaccurately) determined
cosmologically allowed region of parameter space.

We have performed an extensive search of the five-dimensional parameter space
of the model. The five parameters are: the top-quark mass ($m_t$), the
ratio of Higgs vacuum expectation values ($\tan\beta$), and three universal
soft-supersymmetry breaking terms (the gaugino mass $m_{1/2}$, the scalar
mass $m_0$, and the trilinear scalar coupling $A$). The sign of the Higgs
mixing parameter $\mu$ is also undetermined.  Our search accepts only those
points which give adequate radiative electroweak symmetry breaking and
satisfy all known phenomenological constraints in the sparticle and Higgs
spectrum, as described in Ref. \aspects. We also include the very restrictive
proton decay constraint \refs{\AN,\LNP} with the unification scale calculated
using two-loop gauge coupling unification including the effect of light
supersymmetric thresholds \LNPZ. The remaining points in parameter space
($\sim2000$ per sign of $\mu$), which have $\tan\beta=1.5,1.75,2.0$, were then
used to compute the neutralino relic density using the methods of Ref. \LNYa\
for solving the Boltzmann equation and the accurate thermal averaging procedure
of Ref. \GG, as described in Sec. 2.

Since in this model $m_\chi\lsim65\GeV$ (and even $m_\chi\lsim50\GeV$ if
improved lower bounds on $m_h$ are further imposed \LNPWZh), and the Higgs
masses obey $m_h\lsim100\GeV$, $m_{A,H}>500\GeV$, only the $h$ and $Z$
$s$-channel resonances contribute to $\chi\chi\to f\bar f$, with $f$ a light
quark or lepton.\foot{In our calculations, the $h$-width has been obtained
for every point in parameter space including the $h\to f\bar f$ and
$h\to\chi^0_1\chi^0_1$ (when kinematically allowed) channels.}
In addition, the channel $\chi\chi\to hh$ can be kinematically allowed (showing
a threshold effect) in some regions of parameter space ($m_\chi>m_h$). Note
that in the present approach to calculating $\vev{\sigma
v_{\mol}}$, in principle {\it all} kinematically allowed channels contribute,
since $\sqrt{s}\ge2m_\chi$ is allowed. This is in contrast with the traditional
approach where $\sqrt{s}=2m_\chi$ is fixed and only two-body channels with
$m_1+m_2\le2m_\chi$ contribute. However, the farther away $\sqrt{s}$ is from
its lower limit of integration in Eq. \D, the least it will contribute to
the total integral (see Eq. \H{} where $\sqrt{y}=\sqrt{s}/2m_\chi$). This is
exemplified in Fig. 4 which shows that for $y\gsim2$, the contribution to the
integral is negligible. Important annihilation channels that one would need
to worry about (besides $\chi\chi\to f\bar f,hh$) include $\chi\chi\to
W^+W^-,ZZ$, which have $h$-pole and $Z$-pole (only $W^+W^-$) contributions when
the neutralino is not a pure gauge eigenstate. For $m_\chi\lsim50\,(65)\GeV$,
these channels open up for $y\gsim2.6,3.4\,(1.55,2.0)$ respectively. Thus, only
the $W^+W^-$ channel (which requires $\sqrt{s}>2M_W$) could be relevant. To
give any significant contributions, \ie,  $y=s/4m^2_\chi\lsim2$, one would need
$m_\chi\gsim57\GeV$, but in this case $y_Z<0.65$ and
$y_h<(m_h/114\GeV)^2\lsim0.77$ (for $m_h\lsim100\GeV$), and
the poles are not encountered. Therefore, in the following we have only
included the $\chi\chi\to f\bar f,hh$ annihilation channels. The exact
expressions for $w(s)$ are given in the Appendices, including all interference
terms, which have been typically neglected in previous analyses.  We note that
even near the poles (for $y\lsim3$) the interference terms are generally
smaller than the squared terms, although for large $y$ (when all contributions
are unimportant anyway) the two contributions can be comparable.

The results of our calculations are shown in Fig. 5 and Fig. 6 where
$\vev{\sigma v_{\mol}}$ has been computed exactly and in the series expansion
to first order ($a+bx$), respectively. The overall result is not very clear
from these figures, although some shifts of the points are evident. In
particular, around the $Z$-pole ($m_\chi\approx{1\over2}M_Z$) one can see
that the pole structure in the exact case is broader, shallower, and
asymmetric (lower below the pole) relative to the approximate solution,
as anticipated in Sec. 2. Moreover, there is a $\approx53\%\,(27\%)$ increase
in the number of cosmologically allowed points ($\Omega_\chi h^2_0<1$) for
$\mu>0\,(\mu<0)$ relative to the approximate result. This shift is however
not qualitatively significant. In fact, the distribution of cosmologically
allowed points in the $(m_{\chi^\pm_1},m_h)$ plane (see Fig. 7) is very close
to that obtained previously using the series expansion (see Fig. 2 in Ref.
\LNPZ).

To show the effect on the relic density of poles and thresholds of the
annihilation amplitude, we show in Fig. 8 in the $(m_\chi,m_h)$ plane
those points in parameter space where
$\Omega_\chi^{exact}/\Omega_\chi^{approx}>1$. These points correspond to
$\vev{\sigma v_{\mol}}_{exact}/\vev{\sigma v_{\mol}}_{approx}<1$, and according
to Sec. 2 (see Fig. 3) should occur very {\it near} and {\it above} the poles
($y_{Z,h}\lsim1$). This is precisely what Fig. 8 shows, where the $Z$-pole
($m_\chi\approx{1\over2}M_Z$) and the $h$-pole ($m_\chi\approx{1\over2}m_h$)
are denoted by dashed lines. In Fig. 9 we show those points where
$\Omega_\chi^{exact}/\Omega_\chi^{approx}<0.5$, which correspond to
$\vev{\sigma v_{\mol}}_{exact}/\vev{\sigma v_{\mol}}_{approx}\gsim1$, and
according to Fig. 3 should occur {\it below} the poles ($y_{Z,h}\gsim1$). This
is again borne out by the results in Fig. 9. Moreover, for $m_\chi\approx m_h$
the  $\chi\chi\to hh$ channel opens up and a threshold effect causes a drop in
$(\Omega_\chi h^2_0)_{exact}$. These points show up in Fig. 9 along the
diagonal. It is interesting to note that these points lie sligthly {\it above}
the diagonal, \ie, $m_\chi\lsim m_h$. In the usual series approximation,
this channel would not open up until $m_\chi>m_h$. In the exact treatment,
even for $m_\chi\lsim m_h$, $\sqrt{s}\ge2m_h$ is possible and the channel
becomes kinematically allowed. However, as discussed above, only values of
$\sqrt{s}$ close to its minimum value ($\sqrt{s}=2m_\chi$) can contribute
significantly, and this is why the effect only occurs very near the diagonal.
For $m_\chi\ge m_h$ both methods give similar results.

\newsec{Conclusions}
We have investigated the question of the proper thermal averaging of
neutralino annihilation amplitudes which possess poles and thresholds,
following the methods of Ref. \GG. We have focused on two typical resonances
in supersymmetric models, namely the $Z$ boson and the lightest Higgs boson
($h$). In the context of supergravity models with radiative electroweak
symmetry breaking, an exploration of the whole parameter space of the model
is possible and the overall relevance of these sophisticated analyses can
be ascertained. As an example we chose the minimal $SU(5)$ supergravity
model since the presence of such poles is essential to obtain a cosmologically
acceptable model. We have found that the proper thermal averaging is important
for individual points in parameter space. Also, the fraction of cosmologically
acceptable points  is increased somewhat by the accurate procedure. However,
qualitatively the new set of points is very similar to the previously allowed
set.\foot{Concurrently with our calculation there has appeared an analogous
one \ANcosmo\ which reaches similar qualitative conclusions. More quantitative
comparisons are not possible given the lack of detail in Ref. \ANcosmo.}
We conclude that all phenomenological analyses based on the previously
determined cosmologically allowed set remain valid.

\bigskip
\bigskip
\noindent{\it Acknowledgments}: This work has been supported in part by DOE
grant DE-FG05-91-ER-40633. The work of J.L. has been supported  by an SSC
Fellowship. The work of D.V.N. has been supported in part by a grant from
Conoco Inc. The work of K.Y. has been supported by a World-Laboratory
Fellowship. We would like to  thank the HARC Supercomputer Center for the
use of their NEC SX-3 supercomputer.
\vfill\eject

\appendix{A}{Function $w(s)$}
In this Appendix we present the full set of explicit expressions for
the Lorentz invariant function $w(s)$ that we used in this work.
In what follows, all the couplings except those specified otherwise
can be found in Ref. \LNYa.
As discussed in Sec. 3, in the minimal $SU(5)$ supergravity model,
the only relevant annihilation channels are
$\chi\chi\rightarrow{\bar f}f$ and $\chi\chi\rightarrow hh$,
therefore, $w(s)$ can be written as
\eqn\Ai
{w(s)={1\over 32\pi}\Biggl\{\sum_f{c_f\theta (s-4m_f^2)
\sqrt{1-{4m_f^2\over s}}{\wt w}^{({\bar f}f)}(s)}
+\theta (s-4m_h^2)\sqrt{1-{4m_h^2\over s}}
{\wt w}^{(hh)}(s)\Biggr\},}
where $c_f$ is the color factor of conventional fermion $f$ ($c_f=3$
for quarks and $c_f=1$ for leptons), the summation in the
first term runs over all conventional quarks and leptons except
the top quark.
\smallskip
\leftline{A.1. $\chi\chi\rightarrow{\bar f}f$ annihilation}\nobreak

There are totally eight types of contributions
to ${\wt w}^{({\bar f}f)}(s)$:

\noindent{(1) $Z$-boson exchange:}
\eqna\Aii
$$\eqalignno{{\wt w}_Z^{({\bar f}f)}(s)=&{4\over 3}
{|G_A^{\chi\chi Z}|^2\over {(s-M_Z^2)^2+{\Gamma}_Z^2M_Z^2}}
\Biggl\{12|G_A^{ffZ}|^2{m_\chi^2m_f^2\over M_Z^4}(s-M_Z^2)^2\cr
&+\left[|G_A^{ffZ}|^2(s-4m_f^2)+|G_V^{ffZ}|^2(s+2m_f^2)\right]
(s-4m_\chi^2)\Biggr\};&\Aii{}\cr}$$

\noindent{(2) CP-even Higgs-boson ($S_1=h,S_2=H$) exchange:}
\eqn\Aiii
{{\wt w}_S^{({\bar f}f)}(s)=\Biggl|\sum_{i=1,2}
{G_S^{\chi\chi S_i}G_S^{ffS_i}\over
{s-m_{S_i}^2+i{\Gamma}_{S_i}m_{S_i}}}\Biggr|^2(s-4m_f^2)(s-4m_\chi^2);}

\noindent{(3) CP-odd Higgs-boson ($A$) exchange:}
\eqn\Aiv
{{\wt w}_P^{({\bar f}f)}(s)=\Biggl|
{G_P^{\chi\chi A}G_P^{ffA}\over
{s-m_A^2+i{\Gamma}_Am_A}}\Biggr|^2s^2;}

\noindent{(4) $Z$-$A$ interference:}
\eqn\Av
{{\wt w}_{ZP}^{({\bar f}f)}(s)={8\ep m_\chi m_f\over
M_Z^2}Re\left[\left({G_P^{\chi\chi A}G_P^{ffA}\over
{s-m_A^2+i{\Gamma}_Am_A}}\right)
\left({G_A^{\chi\chi Z}G_A^{ffZ}\over
{s-M_Z^2+i{\Gamma}_ZM_Z}}\right)^{*}\right](M_Z^2-s)s;}

\noindent{(5) sfermion (${\wt f}$) exchange:}
\eqna\Avi
$$\eqalignno{
{\wt w}_{\wt f}^{({\bar f}f)}(s)={1\over 4}
\sum_{i,j}\biggl\{
&(A_{+}^iA_{+}^j+B_{+}^iB_{+}^j)\Bigl[
{\cal T}_2-2(m_\chi^2+m_f^2){\cal T}_1+
\bigl((m_\chi^2+m_f^2)^2+4m_\chi^2m_f^2\bigr){\cal T}_0\Bigr]\cr
+&(A_{-}^iA_{-}^j-B_{-}^iB_{-}^j)\Bigl[
{\cal T}_2-2(m_\chi^2+m_f^2){\cal T}_1+(m_\chi^2-m_f^2)^2{\cal T}_0\Bigr]\cr
+&A_{+}^iA_{+}^j\Bigl[
(m_\chi^2+m_f^2)s-4m_\chi^2m_f^2\Bigr]{\cal Y}_1
+A_{-}^iA_{-}^j(m_\chi^2-m_f^2)s{\cal Y}_1\cr
+&(B_{+}^iB_{+}^j-B_{-}^iB_{-}^j)\Bigl[
{\cal Y}_2+(m_\chi^2+m_f^2)^2{\cal Y}_1\Bigr]\cr
+&(A_{+}^iB_{+}^j+A_{+}^jB_{+}^i){\ep}{m_\chi}m_f\Bigl[
4(m_\chi^2+m_f^2){\cal T}_0-4{\cal T}_1+s{\cal Y}_1\Bigr]\cr
-&(A_{+}^iB_{+}^j-A_{+}^jB_{+}^i){\ep}{m_\chi}m_f
{\cal Y}_0\biggl\}(s,m_\chi^2,m_f^2,m_{{\wt f}_i}^2,
m_{{\wt f}_j}^2);&\Avi{}\cr}$$
where ${\cal T}_2$, ${\cal T}_1$, ${\cal T}_0$,
${\cal Y}_2$, ${\cal Y}_1$, ${\cal Y}_0$ are some auxiliary functions
which are given in Appendix B, and the summation runs over two
sfermion mass eigenstates $\wt f_{1,2}$ for each corresponding fermion
$f$. Also, we have introduced the following notation for sfermion
couplings:
\eqna\XXX
$$\eqalignno{A_{\pm}^i\equiv &G_L^i(G_L^i)^{*}\pm
                              G_R^i(G_R^i)^{*},&\XXX a\cr
B_{\pm}^i\equiv &G_L^i(G_R^i)^{*}\pm G_R^i(G_L^i)^{*},&\XXX b\cr}$$
with
\eqna\YYY
$$\eqalignno{G_L^i=&G_L^{\chi f \wt f_L}U_{1i}
                  +G_L^{\chi f \wt f_R}U_{2i},&\YYY a\cr
G_R^i=&G_R^{\chi f \wt f_L}U_{1i}+G_R^{\chi f \wt f_R}U_{2i},&\YYY b\cr}$$
where $U$ is the orthogonal matrix which rotates $\wt f_{L,R}$ into
$\wt f_{1,2}$.

\noindent{(6) $Z$-${\wt f}$ interference:}
\eqn\Avii
{{\wt w}_{Z{\wt f}}^{({\bar f}f)}(s)=
Re\left({G_A^{\chi\chi Z}G_A^{ffZ}\over
{s-M_Z^2+i{\Gamma}_ZM_Z}}\right)I_{Z{\wt f}}^{(1)}
+Re\left({G_A^{\chi\chi Z}G_V^{ffZ}\over
{s-M_Z^2+i{\Gamma}_ZM_Z}}\right)I_{Z{\wt f}}^{(2)}}
where
\eqna\Aviia
$$\eqalignno{I_{Z{\wt f}}^{(1)}=&\sum_i
\biggl\{
4B_{+}^i{\ep}{m_\chi}m_f\Bigl[({s\over M_Z^2}-3)
-\bigl(s+({s\over M_Z^2}-3)(m_\chi^2+m_f^2-m_{{\wt f}_i}^2)\bigr)
{\cal F}(s,m_\chi^2,m_f^2,m_{{\wt f}_i}^2)\Bigr]\cr
-&A_{+}^i\Bigl[
\bigl(2(m_\chi^2+m_f^2)s+{8m_\chi^2m_f^2\over M_Z^2}s
-2(m_\chi^2+m_f^2-m_{{\wt f}_i}^2)^2-16m_\chi^2m_f^2\bigr)
{\cal F}(s,m_\chi^2,m_f^2,m_{{\wt f}_i}^2)\cr
&\qquad\qquad +s+2(m_\chi^2+m_f^2-m_{{\wt f}_i}^2)\Bigr]
\biggr\},&\Aviia a\cr
I_{Z{\wt f}}^{(2)}=&\sum_i
A_{-}^i\Bigl[
\bigl(2(m_\chi^2-m_f^2)s
-2(m_\chi^2+m_f^2-m_{{\wt f}_i}^2)^2
+8m_\chi^2m_f^2\bigr)
{\cal F}(s,m_\chi^2,m_f^2,m_{{\wt f}_i}^2)\cr
&\qquad\qquad +s+2(m_\chi^2+m_f^2-m_{{\wt f}_i}^2)
\Bigr].&\Aviia b\cr}$$

\noindent{(7) $S$-${\wt f}$ interference:}
\eqn\Aviii
{{\wt w}_{S{\wt f}}^{({\bar f}f)}(s)=
\sum_{j=1,2}Re\left({G_S^{\chi\chi {S_j}}G_S^{ff{S_j}}\over
{s-m_{S_j}^2+i{\Gamma}_{S_j}m_{S_j}}}\right)I_{S{\wt f}},}
where
\eqna\Aviiia
$$\eqalignno{I_{S{\wt f}}=\sum_i\biggl\{
&-2A_{+}^i{\ep}{m_\chi}m_f\Bigl[2
+\bigl(s-2(m_\chi^2+m_f^2-m_{{\wt f}_i}^2)\bigr)
{\cal F}(s,m_\chi^2,m_f^2,m_{{\wt f}_i}^2)
\Bigr]\cr
-&B_{+}^i\Bigl[s+\bigl((m_\chi^2+m_f^2-m_{{\wt f}_i}^2)s
-8m_\chi^2m_f^2\bigr){\cal F}(s,m_\chi^2,m_f^2,m_{{\wt f}_i}^2)\Bigr]
\biggr\}.&\Aviiia{}\cr}$$

\noindent{(8) $A$-${\wt f}$ interference:}
\eqn\Aix
{{\wt w}_{A{\wt f}}^{({\bar f}f)}(s)=
Re\left({G_P^{\chi\chi A}G_P^{ffA}\over
{s-m_A^2+i{\Gamma}_Am_A}}\right)I_{A{\wt f}},}
where
\eqna\Aixa
$$\eqalignno{I_{A{\wt f}}=\sum_i\biggl\{
&2A_{+}^i{\ep}{m_\chi}m_f
s{\cal F}(s,m_\chi^2,m_f^2,m_{{\wt f}_i}^2)\cr
+&B_{+}^i\Bigl[-s+(m_\chi^2+m_f^2-m_{{\wt f}_i}^2)
s{\cal F}(s,m_\chi^2,m_f^2,m_{{\wt f}_i}^2)\Bigr]
\biggr\}.&\Aixa{}\cr}$$

\smallskip
\leftline{A.2. $\chi\chi\rightarrow hh$ annihilation}\nobreak

There are totally three types of contributions
to ${\wt w}^{(hh)}(s)$:

\noindent{(1) CP-even Higgs-boson ($S_1=h,S_2=H$) exchange:}
\eqn\Ax
{{\wt w}_S^{(hh)}(s)={1\over 2}\Biggl|\sum_{i=1,2}
{G_S^{\chi\chi S_i}G^{hhS_i}\over
{s-m_{S_i}^2+i{\Gamma}_{S_i}m_{S_i}}}\Biggr|^2(s-4m_\chi^2);}

\noindent{(2) neutralino exchange:}
\eqna\Axi
$$\eqalignno{
{\wt w}_\chi^{(hh)}(s)=&-\sum_{i,j}
|G_S^{\chi{\chi}_ih}|^2|G_S^{\chi{\chi}_jh}|^2
\biggl\{
{\cal T}_2+\Bigl[s+2(m_\chi^2-m_h^2)+2{\ep}m_\chi
({\ep}_im_{{\chi}_i}+{\ep}_jm_{{\chi}_j})\Bigr]{\cal T}_1\cr
-&\Bigl[{\ep}_i{\ep}_jm_{{\chi}_i}m_{{\chi}_j}(s-4m_\chi^2)
-(m_\chi^2-m_h^2)\left(m_\chi^2-m_h^2
+2{\ep}m_\chi({\ep}_im_{{\chi}_i}+{\ep}_jm_{{\chi}_j})\right)
\Bigr]{\cal T}_0\cr
+&\Bigl[\left({\ep}_i{\ep}_jm_{{\chi}_i}m_{{\chi}_j}+
{\ep}m_\chi({\ep}_im_{{\chi}_i}+{\ep}_jm_{{\chi}_j})\right)
(s-4m_\chi^2)-(m_\chi^2-m_h^2)(3m_\chi^2+m_h^2)\Bigr]{\cal Y}_1\cr
+&{\cal Y}_2+{\ep}m_\chi({\ep}_im_{{\chi}_i}-{\ep}_jm_{{\chi}_j}){\cal Y}_0
\biggr\}
(s,m_\chi^2,m_h^2,m_{{\chi}_i}^2,m_{{\chi}_j}^2);&\Axi{}\cr}$$

\noindent{(3) $S$-$\chi$ inteference:}
\eqn\Axii
{{\wt w}_{S\chi}^{(hh)}(s)=2\sum_{j=1,2}
Re\left({G_S^{\chi\chi S_j}G^{hhS_j}\over
{s-m_{S_j}^2+i{\Gamma}_{S_j}m_{S_j}}}\right)I_{S\chi},}
where
\eqna\Axiia
$$\eqalignno{I_{S\chi}=\sum_i|G_S^{\chi{\chi}_ih}|^2\biggl\{
&2{\ep}m_\chi\Bigl[1+(m_\chi^2+m_{{\chi}_i}^2-m_h^2){\cal F}
(s,m_\chi^2,m_h^2,m_{{\chi}_i})\Bigr]\cr
&-{\ep}_im_{{\chi}_i}(s-4m_\chi^2){\cal F}
(s,m_\chi^2,m_h^2,m_{{\chi}_i})\biggr\};&\Axiia{}\cr}$$

\appendix{B}{Some auxiliary functions}
In this Appendix we collect various auxiliary functions
that appear in Appendix A.

First, we define
\eqn\Bi
{t_{\pm}(s,x,y)\equiv x+y-{1\over 2}s\pm {1\over 2}
\sqrt{(s-4x)(s-4y)}}
then
\eqn\Bii
{{\cal F}(s,x,y,z)={1\over \sqrt{(s-4x)(s-4y)}}\ln{\left|
{{t_{+}(s,x,y)-z}\over {t_{-}(s,x,y)-z}}\right|}}
and
\eqna\BB
$$\eqalignno{{\cal T}_2(s,x,y,z_1,z_2)&=1+{1\over (z_1-z_2)}
\Bigl[z_1^2{\cal F}(s,x,y,z_1)-z_2^2{\cal F}(s,x,y,z_2)\Bigr];
&\BB a\cr
{\cal T}_1(s,x,y,z_1,z_2)&={1\over (z_1-z_2)}
\Bigl[z_1{\cal F}(s,x,y,z_1)-z_2{\cal F}(s,x,y,z_2)\Bigr];
&\BB b\cr
{\cal T}_0(s,x,y,z_1,z_2)&={1\over (z_1-z_2)}
\Bigl[{\cal F}(s,x,y,z_1)-{\cal F}(s,x,y,z_2)\Bigr];
&\BB c\cr
{\cal Y}_2(s,x,y,z_1,z_2)&=1+{1\over (s-2x-2y+z_1+z_2)}
\Bigl[z_1(s+z_1-2x-2y){\cal F}(s,x,y,z_1)\cr
&\qquad\qquad\quad +z_2(s+z_2-2x-2y){\cal F}(s,x,y,z_2)\Bigr];
&\BB d\cr
{\cal Y}_1(s,x,y,z_1,z_2)&={1\over (s-2x-2y+z_1+z_2)}
\Bigl[{\cal F}(s,x,y,z_1)+{\cal F}(s,x,y,z_2)\Bigr];
&\BB e\cr
{\cal Y}_0(s,x,y,z_1,z_2)&={1\over (s-2x-2y+z_1+z_2)}
\Bigl[\left(s-2(x+y-z_1)\right){\cal F}(s,x,y,z_1)\cr
&\qquad\qquad\quad -\left(s-2(x+y-z_2)\right){\cal F}(s,x,y,z_2)\Bigr].
&\BB f\cr}$$

\listrefs
\listfigs
\bye